# MIMO Beamforming in Millimeter-Wave Directional Wi-Fi

Keang-Po Ho, *Senior Member, IEEE*, Shi Cheng, *Member, IEEE* and Jianhan Liu, *Member, IEEE*

*Abstract*—Beamforming is indispensable in the operation of 60-GHz millimeter-wave directional multi-gigabit Wi-Fi. Simple power method and its extensions enable the transmitting and receiving antenna arrays to form a beam for single spatial stream. To further improve the spectral efficiency in future 60-GHz directional Wi-Fi, alternating least square (ALS) algorithm can form multiple beams between the transmitter and receiver for multi-input-multi-output (MIMO) operations. For both shared and split MIMO architecture, the ALS beamforming algorithm can be operated in both frequency-flat and frequency-selective channels. In the split architecture, MIMO beamforming approximately maximizes the capacity of the beam-formed MIMO channel.

## I. INTRODUCTION

Unlicensed 60-GHz millimeter-wave band is very suitable for in-room wireless transmissions. With high wall loss, the 60-GHz signal is confined inside a room, reducing interference with neighboring systems using the same frequency band [1][2]. With slight different regional constraints, 60-GHz unlicensed band is almost available worldwide. The channel allocation is also the same in almost all standards, including WirelessHD for video area networking (VAN) [3], IEEE 802.15.3c for personal area network (PAN) [4], and IEEE 802.11ad for wireless local area network (WLAN, or commonly referred to as Wi-Fi) [5]. Having a channel separation of 2160 MHz, the channel bandwidth is sufficient to provide multi-gigabit single-input-single-output (SISO) wireless transmission. Multiple beams can also be formed to provide multi-input-multi-output (MIMO) operations.

SISO operations for 60-GHz Wi-Fi have been defined in the IEEE 802.11ad, once called Wireless Gigabit Alliance (WiGig), as an extension of Wi-Fi from below 6-GHz to 60 GHz. In addition, SISO operations are defined in both WirelessHD [3] and IEEE 802.15.3c [4]. As explained later, beamforming is required in 60-GHz millimeter wave, mostly because of the free-space path loss due to the limited effective antenna aperture in millimeter wavelength. When antenna arrays are used in both transmitter and receiver, as shown in Figure 1, the whole array with many antennas are used to transmit one SISO data stream. Analog phase shifters are used to form a single spatial beam, basically pointing the transmitter and receiver toward each other, providing multi-gigabit yet directional transmission. SISO beamforming based on the simple power method is well-known and can be very effective for frequency-flat channel. Beamforming in frequency-selective channel is less well-known but can be performed using alternating least square (ALS) method that finds iteratively the transmitter beamforming vector by fixing that of the receiver, and then the other way around.

SISO operation for 60-GHz millimeter wave is a mature technology, defined and operated in all 60-GHz standards. The future of 60-GHz Wi-Fi requires MIMO operation to support multiple independent data streams and enhance the spectral efficiency. Ideally, the channel throughput can be increased proportional to the number of spatial streams. Existing SISO 60-GHz Wi-Fi can support up about 7 Gb/s, the future MIMO 60-GHz Wi-Fi should be able to support up to 28 Gb/s in a $4 \times 4$ MIMO beamforming architecture [3]. With many challenging problems to solve, this paper explores the beamforming issues, focusing on the MIMO beamforming algorithm using antenna arrays.

MIMO operation of antenna arrays tries to form multiple beams to support multiple independent data streams using only analog phase shifters. The phase shifters are operated with each set of antennas to enhance the SNR for each spatial stream. In the shared MIMO beamforming architecture, each antenna is driven by the signals from multiple data streams and the beamforming algorithm is just a simple extension from the power method or the ALS method for SISO beamforming.

In the split MIMO architecture, each data stream drives a different set of antennas and each antenna is driven only by one data stream. Also based on the ALS method, the MIMO beamforming algorithm maximizes the signal strength and minimizes interference at the same time. For simplicity, only the algorithm for $2 \times 2$ MIMO beamforming is presented in detail but it can be extended to any $K \times K$ split MIMO beamforming cases, where $K$ is the number of spatial channels.

The remaining parts of this paper are organized as following: Sec. II briefly explains why beamforming is required for 60-GHz Wi-Fi, and defines the mathematical models and notations for later sections. Sec. III describes the SISO beamforming algorithm, especially those for frequency-selective channels. Sec. IV presents the MIMO beamforming algorithm, first for the shared architecture and later for the split architecture. Secs. V and VI are discussion and conclusion, respectively.

## II. BEAMFORMING REQUIREMENT FOR 60-GHZ WI-FI

In this section, the requirement for beamforming in 60-GHz Wi-Fi is first explained. Afterward, the beamforming problem is expressed by its mathematical models, both for frequency-flat and frequency-selective channels, representing as channel matrix and tensor, respectively. This section also defines the notations that are used in later sections.

### A. Antenna Array for Millimeter Wave

According to the Friis' formula for antenna transmission, the received power is proportional to the square of the wavelength. Physically, the received power is always proportional to the effective aperture of the receiving antenna that is proportional to the square of the wavelength. Because of this physical limitation for free space path-loss, compared with typical below 6-GHz Wi-Fi signals, 60-GHz millimeter wave signal is about 28 or 20 dB weaker (calculated using 2.45 and 5.8 GHz) for the same distance. However, due to small wavelength of 60-GHz millimeter wave, many antennas can be arranged as an array and packed into a small package to provide beamforming gain. Figure 1 shows the schematic diagram of a typical 60-GHz transmitter and receiver using antenna arrays. The antenna arrays can be used to compensate for the transmission loss if proper beamforming is applied using the phase shifters in Figure 1.

In Figure 1, the transmitting signal splits to many power amplifiers (PAs) and antennas. When a transmitting signal in different antennas is phase shifted by phase shifters, the signal can be steered to different direction, constructively enhanced in certain direction, and destructively nulled in other directions [6][7]. Using phase shifters in different antennas, the receiving antenna array can also enhance the signal in certain direction and null the signal in other directions.

Comparing the 802.11ad using 60 GHz with 802.11ac using 5 GHz, in additional to free space loss, 802.11ad based 60-GHz Wi-Fi also requires different power. Specially, a 4.6 Gb/s single-carrier signal using 16-quadrature amplitude modulation (QAM) in 60-GHz 802.11ad may compare with 780 Mb/s/antenna 256-QAM signal in 160-MHz 5-GHz 802.11ac, both using the same coding rate of 3/4 and operating to a distance of about 10 m (reasonable for both cases). The signal bandwidth, and thus the noise, has about 10.7 dB difference (1.76 GHz versus 151 MHz). The modulation scheme, 16- compared with 256-QAM, gives a difference of -12.3 dB in required signal-to-noise ratio (SNR). Combined together, 60-GHz signal actually requires about 1.6 dB less power than 5-GHz signal.

Compared with below 6-GHz Wi-Fi, the poor efficiency for 60-GHz millimeter wave PA and the high noise figure for the corresponding low-noise amplifier (LNA) are both not helpful for 802.11ad 60-GHz Wi-Fi. Because of the PA efficiency and LNA noise, 60-GHz Wi-Fi may require 6-7 dB larger transmitting power than 5-GHz Wi-Fi. Combining the free space loss, signal characteristic, and circuitry limitation, 60-GHz Wi-Fi may require an addition gain of about 25 dB compared with 5-GHz Wi-Fi.

Beamforming using an antenna array is essential to bridge the gap due to channel loss between 60-GHz and below 6-GHz Wi-Fi. Ideally, $N$ transmitting antennas can provide a beamforming gain of $N$ given the same total transmitting power, and the total gain becomes $N^2$ because $N$ PAs emit $N$ times more power. At the same time, $M$ receiving antennas can provide a SNR gain of $M$. The total ideal gain provided by two $N \times M$ antenna arrays is $20\log_{10}N + 10\log_{10}M$ in decibel unit. For example, for antenna arrays with $N = M = 8$ to 16 ($8 \times 8$ to $16 \times 16$) can provide an ideal beamforming gain of 27 to 36 dB.

In 60-GHz SISO beamforming, multiple antennas use the same number of PAs as shown in Figure 1, increase the power consumption by a factor of $N$ without corresponding increases in data rate. The increase in data rate in 60-GHz Wi-Fi is mostly due to the availability of bandwidth and beamforming provides SNR improvement. In 5-GHz MIMO Wi-Fi, the number of independent data streams is ideally proportional to the minimum of transmitting and receiving antennas, increasing the data rate proportionally. Technically, multiple antennas in 60 GHz using SISO beamforming can only provide diversity and SNR gain but multiple antennas in 5 GHz may ideally provide multiplexing gain (diversity gain as well a choice [8]).

MIMO beamforming in 60-GHz Wi-Fi is essential to further increase the system throughput. Figure 2 shows the measurement of the impulse response energy profile between two $36 \times 36$ antenna arrays. Each impulse corresponds to different reflectors that can support different beams. With up to 7 reflections within 2-3 dB, the antenna arrays definitely can support MIMO operations and the remaining question is to find a method to utilize them.

### B. Beamforming Model and Notations

The SISO beamforming model is the simplest for frequency-flat channel with a time (or frequency) independent $N \times M$ channel matrix $\mathbf{H}$ that can be processed using methods from matrix analysis [9]. The optimal beamforming is to find the two complex column vectors for transmitting $\mathbf{u}$ ($N$ elements) and receiving $\mathbf{v}$ ($M$ elements) to maximize $\|\mathbf{u}^H\mathbf{H}\mathbf{v}\|$, where $^H$ denotes the Hermitian transpose and both $\mathbf{u}$ and $\mathbf{v}$ has unity norm of $\|\mathbf{u}\| = \|\mathbf{v}\| = 1$. Those vectors are applied to the transmitting and receiving phase shifters of Figure 1, may be together with adjustment of PA and LNA gain. The SISO beamforming for frequency flat channel is very simple by using the power method.

The profile of Figure 2 shows that typical impulse response is not frequency-flat as single impulse in time-domain, but characterized by multiple discrete impulses.

For frequency-selective channel, the channel matrix is either time or frequency dependent as shown in Figure 2. With discrete time sampling, the channel may be expressed as a tensor $\mathcal{H}$ that is defined as three-way signal [10][11] instead of physical tensor product. The tensor $\mathcal{H}$ has elements as $h_{n,m,p}$, where $n$ from 1 to $N$ is the index of the transmitting antenna, $m$ from 1 to $M$ is the index of the receiving antenna, and $p$ is an index for either time or frequency. As the impulse response of wireless channel is typically some discrete pulses, expressing $p$ as time index is more convenient and is used in later parts of this paper. However, the model here can also be used when $p$ is an index for frequency. The energy profile of Figure 2 is basically $\sum_n \sum_m |h_{n,m,p}|^2$ versus the timing index $p$. We will not define the number samples for time index $p$. Practically, for beamforming purpose, the algorithm may select all timing indices as shown in Figure 2, or select a few peaks. The beamforming algorithm does not assume that the indexes $p$ and $p+1$ are for adjacent timing samples. The strongest peak of Figure 2 is less than 10 dB larger than the measurement floor, showing that incoherent combination of signal does not give good channel gain and beamforming with coherent combination is required.

The notations of this paper are defined as follows. The calligraphic upper case font $\mathcal{H}$ is used to denote a tensor and the bold upper case font $\mathbf{H}$ is for a matrix. Later on, the matrix $\mathbf{H}_p$ denotes the tensor $\mathcal{H}$ at time index $p$, or all the elements of $h_{n,m,p}$ with a fixed $p$. The elements for tensor $\mathcal{H}$ and matrix $\mathbf{H}$ are $h_{n,m,p}$ and $h_{n,m}$, respectively. The elements with subscript indexes are not defined separately after the tensor or matrix is defined. The lower case bold font $\mathbf{u}$ or $\mathbf{v}$ denotes column vector. All beamforming vectors are normalized with unity norm $\|\mathbf{u}\| = \|\mathbf{v}\| = 1$. As shown earlier, the notation $^H$ is the Hermitian transpose with transpose and complex conjugate. The transpose and complex conjugate are denotes as $^T$ and $*$, respectively, for example, $\mathbf{A}^H = \mathbf{A}^{T*}$.

Later on, we may multiply a tensor $\mathcal{H}$ with a matrix $\mathbf{A}$ (or a vector $\mathbf{a}$ as special case). As an example, the elements for $\mathcal{B} = \mathcal{H} \times_2 \mathbf{A}$ are

$$b_{n,m,p} = \sum_l h_{n,l,p} a_{l,m}, \quad (1)$$

where the subscript $_2$ indicates the multiplication index. A dimension with only one element will be dropped and $\mathcal{H} \times_2 \mathbf{a}$ is a matrix.

All phase shifters in Figure 1 are assumed wideband device, for tensor channel, SISO beamforming gives a channel vector of

$$\mathbf{h} = \mathcal{H} \times_1 \mathbf{u}^* \times_2 \mathbf{v} \quad (2)$$

that is equivalent to a finite-impulse-response (FIR) filter. One of the options is to optimize the channel capacity as given by the FIR channel $\mathbf{h}$, using water filling in the frequency domain. Another option may include the possible equalizer and maximize the SNR after equalization. Unfortunately, either the channel capacity or equalized system leads to complicate and difficult optimization issues [12].

The viable yet reasonable objective is to maximize the SNR of the system before equalization, or equivalently, to maximize the norm of $\|\mathbf{h}\|$, as a reminder again, $\|\mathbf{u}\| = \|\mathbf{v}\| = 1$. As shown later, beamforming to maximize system SNR leads to iterative ALS algorithm for the optimal beamforming vectors.

The beamforming algorithm of tensor channel $\mathcal{H}$ is the major concern here. Many implementation issues are important for the system to operate but will not discuss in details here. The beamforming algorithms are verified using $\mathbf{H}$ as random Gaussian matrix or $\mathcal{H}$ as random Gaussian tensor. Although both of them may not correspond to a physical channel, those are very helpful for repeatable, verifiable, and accessible algorithm validation.

### III. SISO BEAMFORMING

SISO beamforming puts up a single beam between the transmitter and receiver that may be the simplest beamforming, especially for frequency-flat channel. This section will first give an overview for the beamforming for matrix channel $\mathbf{H}$ and then two different beamforming algorithms for the tensor channel $\mathcal{H}$. Those algorithms present the fundamental concepts for MIMO beamforming in next section.

#### A. Power Method for Matrix Channel

The beamforming algorithm for frequency-flat channel, represented by the channel matrix $\mathbf{H}$, is well-known. Mathematically, the channel matrix $\mathbf{H}$ has its singular value decomposition (SVD) $\mathbf{H} = \mathbf{U\Sigma V}^H$, where $\mathbf{U}$ and $\mathbf{V}$ are unitary matrices, and $\mathbf{\Sigma}$ is a diagonal matrix with all singular values [9][13]. The transmitting and receiving beamforming vectors are the left $\mathbf{u}$ and right $\mathbf{v}$ eigenvectors that correspond to the largest singular value $\sigma_1$. In another interpretation, those two vectors also give the least square approximation for $\mathbf{H}$ to minimize $\|\mathbf{H} - \sigma_1 \mathbf{uv}^H\|^2$ or maximize the Rayleigh quotient of $\sigma_1 = \|\mathbf{u}^H \mathbf{Hv}\|$.

SISO beamforming does not need to find the full SVD for the channel matrix $\mathbf{H}$. The maximum Rayleigh quotient $\|\mathbf{u}^H \mathbf{Hv}\|$ can be found using the simple power method, as shown in Algorithm I.

ALGORITHM **I**: SIMPLE POWER METHOD
Initial   Pick an initial vector $\mathbf{v}^{(0)}$, $k = 0$
Step 1   Increment $k$, calculate:
   a.   $\tilde{\mathbf{u}}^{(k)} = \mathbf{Hv}^{(k-1)}$, $\mathbf{u}^{(k)} = \tilde{\mathbf{u}}^{(k)} / \|\tilde{\mathbf{u}}^{(k)}\|$
   b.   $\tilde{\mathbf{v}}^{(k)} = \mathbf{H}^H \mathbf{u}^{(k)}$, $\mathbf{v}^{(k)} = \tilde{\mathbf{v}}^{(k)} / \|\tilde{\mathbf{v}}^{(k)}\|$
Step 2   Repeat step 1 until the convergence of $\mathbf{u}^{(k)}$, $\mathbf{v}^{(k)}$, and $\sigma_1 = \|\tilde{\mathbf{v}}^{(k)}\|$.

This algorithm is also called power iteration [9] or Rayleigh quotient iteration. As long as the initial vector $\mathbf{v}^{(0)}$ is not exactly aligned with one of the right eigenvectors (other than $\mathbf{v}$ corresponding to $\sigma_1$) of the channel matrix $\mathbf{H}$, the algorithm converges to the global optimum after some iterations [9]. In practice and due to noise and error, the algorithm always converges.

In step 1a, the simple power method fixes the receiving beamforming vector $\mathbf{v}$ to find the optimal transmitting beamforming vector $\mathbf{u}$. Step 1b is just the other way around for step 1a. This alternating optimization method will be used later for many other cases. The simple power method is also used in laser cavity design [14] and the PageRank algorithm [15]. Both steps 1a and 1b are similar to the methods in [6][7] for receiver beamforming.

### B. ALS for Tensor Channel

Beamforming for tensor channel is to find the beamforming vectors **u** and **v** to maximize the channel power given by $\|\mathbf{h}\|^2$ with **h** from (2). The ALS method tries to find the optimal **u** given a choice of **v**, and the other way around. Given a choice of **v**, we obtain the matrix $\mathbf{A} = \mathcal{H} \times_2 \mathbf{v}$ and the optimal beamforming vector **u** can be obtained by the least square approximation for $\|\mathbf{A} - \sigma_1 \mathbf{u}\mathbf{a}^H\|^2$ with another vector **a** to allocate power to each impulse of the channel tensor $\mathcal{H}$. After obtaining the matrix $\mathbf{A} = \mathcal{H} \times_2 \mathbf{v}$, the optimal vector **u** can be calculated using the simple power method of Algorithm I. Similarly, given **u**, we may obtain the matrix $\mathbf{B} = \mathcal{H} \times_1 \mathbf{u}^*$ and the corresponding least square approximation of $\|\mathbf{B} - \sigma'_1 \mathbf{b}\mathbf{v}^H\|^2$ using Algorithm I. This algorithm is similar to those in [16]-[18] but with significant differences.

**ALGORITHM II**: ALTERNATING LEAST SQUARE (ALS)
Initial   Pick an initial vector $\mathbf{v}^{(0)}$, $k = 0$.
Step 1   Increment $k$:
    a. Calculate $\mathbf{A}^{(k)} = \mathcal{H} \times_2 \mathbf{v}^{(k-1)}$. Find $\mathbf{u}^{(k)}$ to minimize
       $\|\mathbf{A}^{(k)} - \sigma_1^{(k)}\mathbf{u}^{(k)}\mathbf{a}^{(k)H}\|^2$
    b. Calculate $\mathbf{B}^{(k)} = \mathcal{H} \times_1 \mathbf{u}^{(k)*}$. Find $\mathbf{v}^{(k)}$ to minimize
       $\|\mathbf{B}^{(k)} - \sigma'^{(k)}_1 \mathbf{b}^{(k)}\mathbf{v}^{(k)H}\|^2$
Step 2   Repeat step 1 until the convergence of $\mathbf{u}^{(k)}$, $\mathbf{v}^{(k)}$, and $\sigma_1^{(k)}$.

Algorithm II using ALS always converges to a solution regardless of the initial vector $\mathbf{v}^{(0)}$. Unlike the simple power method of Algorithm I, converging to global optimum depends on the initial vector $\mathbf{v}^{(0)}$. It may be very easy to find special case for the algorithm to converge to local minimum. For example, if $\mathcal{H}$ is a $N \times M \times 2$ tensor and the rank of $\mathbf{H}_1$ and $\mathbf{H}_2$ is one, it is easy to find initial vector to converge to a vector closer to the eigenvector of either $\mathbf{H}_1$ or $\mathbf{H}_2$. Numerically, several random initial vectors may be used and the beamforming vectors $\mathbf{u}^{(k)}$ and $\mathbf{v}^{(k)}$ may be chosen for that with the maximum $\sigma_1^{(k)}$. In practical simulation, if the initial $\mathbf{v}^{(0)}$ is chosen as the right eigenvector for the maximum singular value from the matrix $[\mathbf{H}_1^T, \mathbf{H}_2^T, ...]^T$, the beamforming vectors rarely converges to local optimum but counter-example can be found.

The major drawback for the ALS Algorithm II is the required operations to minimize $\|\mathbf{A}^{(k)} - \lambda_1^{(k)}\mathbf{u}^{(k)}\mathbf{a}^{(k)H}\|^2$ and $\|\mathbf{B}^{(k)} - \lambda'^{(k)}_1 \mathbf{b}^{(k)}\mathbf{v}^{(k)H}\|^2$. Algorithm I may be used for the purpose but may need several iterations by itself. In practice, the iteration may speed up by finding the eigenvectors for, as an example, $\mathbf{A}^{(k)}\mathbf{A}^{(k)H} = \sum_p \mathbf{H}_p \mathbf{v}^{(k-1)}\mathbf{v}^{(k-1)H}\mathbf{H}_p^H$.

### C. High-Order Power Method for Tensor Channel

High-order power method (HOPM) of Algorithm III is an extension of the simple power method Algorithm I for rank-1 approximation for a tensor [11]. For tensor channel $\mathcal{H}$, one more step is required in addition to Algorithm I.

**ALGORITHM III**: HIGH-ORDER POWER METHOD (HOPM)
Initial   Pick initial vectors $\mathbf{u}^{(0)}$ and $\mathbf{v}^{(0)}$, $k = 0$
Step 1   Increment $k$, calculate:
    a. $\mathbf{h}^{(k)} = \mathcal{H} \times_1 \mathbf{u}^{(k-1)*} \times_2 \mathbf{v}^{(k-1)}$, $\tilde{\mathbf{h}}^{(k)} = \mathbf{h}^{(k)}/\|\mathbf{h}^{(k)}\|$
    b. $\tilde{\mathbf{u}}^{(k)} = \mathcal{H} \times_2 \mathbf{v}^{(k-1)} \times_3 \tilde{\mathbf{h}}^{(k)*}$, $\mathbf{u}^{(k)} = \tilde{\mathbf{u}}^{(k)}/\|\tilde{\mathbf{u}}^{(k)}\|$
    c. $\tilde{\mathbf{v}}^{(k)} = \mathcal{H} \times_1 \mathbf{u}^{(k)*} \times_3 \tilde{\mathbf{h}}^{(k)*}$, $\mathbf{v}^{(k)} = \tilde{\mathbf{v}}^{(k)}/\|\tilde{\mathbf{v}}^{(k)}\|$
Step 2   Repeat step 1 until the convergence of $\mathbf{u}^{(k)}$, $\mathbf{v}^{(k)}$, and $\sigma_1 = \|\tilde{\mathbf{v}}^{(k)}\|$.

Compared with the simple power method of Algorithm I, the HOPM Algorithm III adds another step to find $\tilde{\mathbf{h}}^{(k)}$ as the best power allocation along different pulses. The vector $\tilde{\mathbf{h}}^{(k)}$ is the same as $\mathbf{a}^{(k)}$ and $\mathbf{b}^{(k)}$ in steps 1a and 1b for the ALS Algorithm II. Instead of finding the optimal vectors for $\mathbf{a}^{(k)}$ and $\mathbf{b}^{(k)}$ by one iteration in Algorithm II, Algorithm III finds the vector $\tilde{\mathbf{h}}^{(k)}$ step by step. Compared to Algorithms II, Algorithm III based on HOPM may require larger number of iterations for convergence.

The HOPM Algorithm III also requires an initial vector similar to the ALS Algorithm II. The vector $\mathbf{v}^{(0)}$ may be chosen as the same as Algorithm II. The vector $\mathbf{u}^{(0)}$ may be obtained similarly from the left eigenvector for the largest singular value from the matrix $[\mathbf{H}_1, \mathbf{H}_2, ...]$.

### D. Numerical Results

Figure 3 shows the simulated distributions for the largest singular value $\sigma_1$ or its equivalent for SISO beamforming. The channel is for arrays with $N = M = 16$ transmitting and receiving antennas. For matrix channel **H**, the largest singular value $\sigma_1$ is found by Algorithm I using simple power method. For tensor channel $\mathcal{H}$, the singular value is given by the converged $\sigma_1$ for ALS Algorithm II and HOPM Algorithm III. All algorithms iterate 8 times to show the effect of convergence. Equivalently, the largest singular value for tensor channel is the norm for the FIR channel $\|\mathbf{h}\|$, given by (2).

As explained earlier, the algorithms are tested using complex Gaussian channel, with each element having unity variance. The tensor channel $\mathcal{H}$ has dimension of $16 \times 16 \times 2$. Each curve in Figure 3 is the results from 200,000 random channels.

Figure 3 shows that tensor channel $\mathcal{H}$ is just slightly better than the matrix channel although the system receives twice the power. Both ALS Algorithm II and HOPM Algorithm III obtain more or less the same singular value. The HOPM Algorithm III obtains slightly smaller singular value, due to slower convergence rate than ALS Algorithm II. If number of iterations is large (100 was tested), both ALS Algorithm II and HOPM Algorithm III converge to the same singular value and beamforming vectors.

## IV. MIMO Beamforming

To further increase the data rate, the antenna arrays must be able to form multiple spatial beams. MIMO beamforming is described in this section.

MIMO beamforming can be implemented using shared or split architecture as shown in Figure 4 and Figure 5, respectively. In both architectures, each data stream uses independent phase shifters. In the shared architecture shown in Figure 4, multiple phase shifters share the same PA and antenna. The phase shifters sharing the same PA are driven individually by independent data streams. In the split architecture shown in Figure 5, each antenna is driven by a single data stream. The same PA and antenna are used only by one phase shifter.

In the shared architecture, the phase shifters in both transmitter and receiver can be used to form orthogonal spatial beams in matrix channel without interference between beams. In practice, MIMO receiver is still required as the beamforming vectors have estimation errors. For tensor channel, the phase shifters cannot always from orthogonal spatial beams even in the shared architecture. For optimal performance, digitally implemented pre-coding matrix may be required in transmitter for tensor channel.

In the split architecture of Figure 5, pre-coding matrix is required in the transmitter for both matrix and tensor channels. Joint MIMO processing is required in the receiver as the spatial beams may interfere with each other.

### A. Shared MIMO Architecture

MIMO beamforming in the shared architecture of Figure 4 is very straight-forward for frequency-flat channel with channel matrix $\mathbf{H}$. In the SVD of $\mathbf{H} = \mathbf{U\Sigma V}^H$, the optimal beamforming vectors are the eigenvectors corresponding to the first $K$ largest singular values of $\sigma_1$ to $\sigma_K$, assuming that $K$ is the number of spatial streams. The algorithm to find those pairs of beamforming vectors [19] is a modification of the simple power method of Algorithm I and will not further elaborate here. Ideally, no MIMO processing is required in the transmitter and receiver, because the beams are orthogonal to each other. In practice, joint MIMO processing is required in the receiver to combat estimation error or channel variations.

For frequency-selective channel represents by a channel tensor $\mathcal{H}$, the ALS Algorithm II can be modified to find $K$ beamforming vectors. Given a choice of $\mathbf{V}$ as a $M \times K$ matrix represent $K$ beamforming vectors, the product $\mathcal{A} = \mathcal{H} \times_2 \mathbf{V}$ is still a tensor. A Hermitian matrix $\mathbf{D} = \sum_p \mathbf{H}_p \mathbf{V}\mathbf{V}^H \mathbf{H}_p^H$ can be found that is similar to $\mathbf{A}^{(k)}\mathbf{A}^{(k)H}$ for Algorithm II. The ALS for shared MIMO can be derived as Algorithm IV.

**ALGORITHM IV**: ALS FOR SHARED MIMO
Initial    Pick $K$ initial vectors for the matrix $\mathbf{V}^{(0)}$, $k = 0$.
Step 1    Increment $k$.
  a.    Calculate $\mathcal{A}^{(k)} = \mathcal{H} \times_2 \mathbf{V}^{(k-1)}$, $\mathbf{D}^{(k)} = \sum_p \mathbf{A}_p^{(k)} \mathbf{A}_p^{(k)H}$. Find $\mathbf{U}^{(k)}$ as the $K$ eigenvectors corresponding to the $K$ largest eigenvalues of $\mathbf{D}^{(k)}$.
  b.    Calculate $\mathcal{B}^{(k)} = \mathcal{H} \times_1 \mathbf{U}^{(k)*}$, $\mathbf{E}^{(k)} = \sum_p \mathbf{B}_p^{(k)T} \mathbf{B}_p^{(k)}$. Find $\mathbf{V}^{(k)}$ as the $K$ eigenvectors corresponding to the $K$ largest eigenvalues for $\mathbf{E}^{(k)}$.
Step 2    Repeat step 1 until the convergence of $\mathbf{U}^{(k)}$ and $\mathbf{V}^{(k)}$.

The ALS Algorithm IV is also converged step by step but may be to a local optimum. Initialized $\mathbf{V}^{(0)}$ by the right eigenvectors corresponding to the $K$ largest singular values for the matrix $\left[\mathbf{H}_1^T, \mathbf{H}_2^T, ...\right]^T$ is helpful. Using $\mathbf{U}$ and $\mathbf{V}$ obtained by the ALS MIMO Algorithm IV, the beamformed channel remains a tensor channel of

$$\mathcal{X} = \mathcal{H} \times_1 \mathbf{U}^* \times_2 \mathbf{V}. \quad (3)$$

Unlike the case with matrix channel $\mathbf{H}$, the tensor channel is not likely to be "diagonal" without interference between different beams. Digital pre-coding matrix is required in the transmitter for optimal performance. The performance for the beamformed tensor channel $\mathcal{X}$ will be discussed further in Sec. IV.C.

The HOPM Algorithm III can also be modified for MIMO beamforming for shared architecture but cannot reduce the computation complexity per iteration [11] and thus are not studied further here.

### B. Split MIMO Architecture for Matrix Channel

In the split architecture of Figure 5, each phase shifter is connected to only one PA and antenna. Each spatial data stream, may be pre-coded by a $K \times K$ unitary matrix, drives different set of phase shifters and sends the signal to the corresponding set of PAs and antennas.

In the simplest case for a $2 \times 2$ MIMO with $N_1$ and $N_2$ transmitting antennas for two spatial data streams, respectively, where $N = N_1 + N_2$, and $M_1$ and $M_2$ receiving antennas for the two spatial data streams, respectively, where $M = M_1 + M_2$. The objective for MIMO beamforming is to maximize the throughput (or other equivalent or approximated objectives) for the following $2 \times 2$ beamformed MIMO matrix channel:

$$\Xi = \begin{bmatrix} \mathbf{u}_1^H \mathbf{H}_{11} \mathbf{v}_1 & \mathbf{u}_1^H \mathbf{H}_{12} \mathbf{v}_2 \\ \mathbf{u}_2^H \mathbf{H}_{21} \mathbf{v}_1 & \mathbf{u}_2^H \mathbf{H}_{22} \mathbf{v}_2 \end{bmatrix}, \quad (4)$$

where $\mathbf{H}_{11}$, $\mathbf{H}_{12}$, $\mathbf{H}_{21}$, and $\mathbf{H}_{22}$ are block matrix with size of $N_1 \times M_1$, $N_1 \times M_2$, $N_2 \times M_1$, and $N_2 \times M_2$, respectively, $\mathbf{u}_1$ and $\mathbf{u}_2$ are column vectors with $N_1$ and $N_2$ elements, respectively, and $\|\mathbf{u}_1\| = \|\mathbf{u}_2\| = 1$, $\mathbf{v}_1$ and $\mathbf{v}_2$ are column vectors with $M_1$ and $M_2$ elements, respectively, and $\|\mathbf{v}_1\| = \|\mathbf{v}_2\| = 1$.

The optimization for the channel matrix $\Xi$ is not very straightforward. Assume that $\Xi$ has two singular values of $\sigma_1$ and $\sigma_2$, with $\sigma_1 \geq \sigma_2$. If the objective is to maximize $\sigma_1$, simple algebra shows that it is approximately the same as Algorithm I using simple power method to find the optimal $\mathbf{u}$ and $\mathbf{v}$, and for example, $\mathbf{u}_1$ is the first $N_1$ elements of $\mathbf{u}$ and normalizes to $\|\mathbf{u}_1\| = 1$ afterward, and $\sigma_2$ is approximately equal to zero.

Therefore, the optimization for the channel matrix $\Xi$ needs to take into account both singular values $\sigma_1$ and $\sigma_2$ together. Another bad choice is to maximize the channel power of $\sigma_1^2 + \sigma_2^2 = \text{Tr}[\Xi\Xi^H]$ because the solution will be $\sigma_2 = 0$ with maximum $\sigma_1$, where $\text{Tr}[\ ]$ is the trace of a matrix.

The feasible and reasonable choice is to maximize the product of $\sigma_1\sigma_2 = |\det[\Xi]|$, where $\det[\ ]$ is the determinant of a matrix. In high SNR, the overall channel capacity is equal to $\log_2(1+\sigma_1^2\chi) + \log_2(1+\sigma_2^2\chi) \approx 2\log_2(\sigma_1\sigma_2) + 2\log_2\chi$, where $\chi$ is the ratio of transmit power to receive noise. In high SNR, maximizing $\sigma_1\sigma_2 = |\det[\Xi]|$ also approximately maximizes the channel throughput because logarithmic function is a monotonic function.

The ALS method can be used to maximize $|\det[\Xi]|$ by the optimization of $\mathbf{u}_1$ and $\mathbf{u}_2$ given $\mathbf{v}_1$ and $\mathbf{v}_2$, and the other way around. In practice, the value of $\det[\Xi]$ can always be a positive real value by changing the phase of both $\mathbf{v}_1$ and $\mathbf{v}_2$ (or $\mathbf{u}_1$ and $\mathbf{u}_2$) together. Given $\mathbf{v}_1$ and $\mathbf{v}_2$, the expression $\det[\Xi]$ can be rewritten as

$$\begin{vmatrix} \mathbf{u}_1^H\mathbf{H}_{11}\mathbf{v}_1 & \mathbf{u}_1^H\mathbf{H}_{12}\mathbf{v}_2 \\ \mathbf{u}_2^H\mathbf{H}_{21}\mathbf{v}_1 & \mathbf{u}_2^H\mathbf{H}_{22}\mathbf{v}_2 \end{vmatrix} = \begin{vmatrix} \mathbf{u}_1^H\mathbf{H}_{11}\mathbf{v}_1 & \mathbf{u}_1^H\mathbf{H}_{12}\mathbf{v}_2 \\ \mathbf{v}_1^T\mathbf{H}_{21}^T\mathbf{u}_2^* & \mathbf{v}_2^T\mathbf{H}_{22}^T\mathbf{u}_2^* \end{vmatrix}, \quad (5)$$

or $\det[\Xi] = \mathbf{u}_1^H\mathbf{A}\mathbf{u}_2^*$ with

$$\mathbf{A} = \mathbf{H}_{11}\mathbf{v}_1\mathbf{v}_2^T\mathbf{H}_{22}^T - \mathbf{H}_{12}\mathbf{v}_2\mathbf{v}_1^T\mathbf{H}_{21}^T, \quad (6)$$

and $\det[\Xi]$ is maximized by $\mathbf{u}_1$ and $\mathbf{u}_2$ by the least square fit of $\|\mathbf{A} - \lambda_1\mathbf{u}_1\mathbf{u}_2^T\|^2$ with $\det[\Xi] = \lambda_1$, similar to the Rayleigh quotient problem for Algorithm I.

Based on the ALS method, the MIMO beamforming for split architecture is Algorithm V.

**ALGORITHM V**: ALS FOR SPLIT MIMO MATRIX CHANNEL

Initial   Pick initial vectors $\mathbf{v}_1^{(0)}$ and $\mathbf{v}_2^{(0)}$, $k=0$.

Step 1   Increment $k$.
 a. Calculate
 $\mathbf{A}^{(k)} = \mathbf{H}_{11}\mathbf{v}_1^{(k-1)}\mathbf{v}_2^{(k-1)T}\mathbf{H}_{22}^T - \mathbf{H}_{12}\mathbf{v}_2^{(k-1)}\mathbf{v}_1^{(k-1)T}\mathbf{H}_{21}^T$. Find $\mathbf{u}_1^{(k)}$ and $\mathbf{u}_2^{(k)}$ to minimize $\|\mathbf{A}^{(k)} - \lambda_1^{(k)}\mathbf{u}_1^{(k)}\mathbf{u}_2^{(k)T}\|^2$.
 b. Calculate $\mathbf{B}^{(k)} = \mathbf{H}_{11}^T\mathbf{u}_1^{(k)*}\mathbf{u}_2^{(k)H}\mathbf{H}_{22} - \mathbf{H}_{21}^T\mathbf{u}_2^{(k)*}\mathbf{u}_1^{(k)H}\mathbf{H}_{21}$. Find $\mathbf{v}_1^{(k)}$ and $\mathbf{v}_2^{(k)}$ to minimize $\|\mathbf{B}^{(k)} - \lambda_1^{(k)}\mathbf{v}_1^{(k)*}\mathbf{v}_2^{(k)H}\|^2$.

Step 2   Repeat step 1 until the convergence of $\mathbf{u}_1^{(k)}$, $\mathbf{u}_2^{(k)}$, $\mathbf{v}_1^{(k)}$, and $\mathbf{v}_2^{(k)}$

Step 3   Calculate $\Xi$ and obtain the corresponding singular values.

Algorithm V converges step by step. Both $\mathbf{A}^{(k)}$ and $\mathbf{B}^{(k)}$ have up to two non-zero singular values. In one interpretation, the singular value for $\mathbf{A}$ (6) maximizes the contribution from both diagonal block matrices $\mathbf{H}_{11}$ and $\mathbf{H}_{22}$ but minimizes the contribution from the off-diagonal interfering block matrices $\mathbf{H}_{12}$ and $\mathbf{H}_{21}$. The system is also equivalent if $\mathbf{A}$ (6) minimizes the contribution from both diagonal block matrices $\mathbf{H}_{11}$ and $\mathbf{H}_{22}$ but maximizes the contribution from the off-diagonal interfering block matrices $\mathbf{H}_{12}$ and $\mathbf{H}_{21}$.

The ALS Algorithm V optimizes the beamforming vectors alternatively in transmitter and receiver. Other variation is also possible, for example, given vectors of $\mathbf{v}_1$ and $\mathbf{u}_1$, the optimal $\mathbf{v}_2$ and $\mathbf{u}_2$ may be obtained by the matrix of

$$\left(\mathbf{u}_1^H\mathbf{H}_{11}\mathbf{v}_1\right)\mathbf{H}_{22} - \mathbf{H}_{21}\mathbf{v}_1\mathbf{u}_1^H\mathbf{H}_{12} \quad (7)$$

using ALS method. However, variations like (7) may be difficult to operate in practice.

The ALS Algorithm V can also divide into more steps. The least square approximation to minimize $\|\mathbf{A} - \lambda_1\mathbf{u}_1\mathbf{u}_2^T\|^2$, for example, may be conducted by the simple power method Algorithm I. The algorithm may first optimize $\mathbf{u}_1$ and then $\mathbf{u}_2$, equivalent to just operate one iteration of Algorithm I.

The ALS Algorithm IV for split MIMO is for $2\times 2$ beamformed MIMO but can be extended to the general case for $K\times K$ MIMO with the objective to maximize the product of the singular values or $\det[\Xi]$. The operation is equivalent to rewrite $\det[\Xi]$ as matrix multiplication according to expressions similar to (6) or (7) but with a total of $K!$ terms.

*C. Split MIMO Architecture for Tensor Channel*

For frequency-selective tensor channel $\mathcal{H}$, beamforming for split MIMO architecture of Figure 5 is not as simple as that for the matrix channel $\mathbf{H}$. Using $2\times 2$ MIMO as example, the $2\times 2$ MIMO channel is

$$\mathcal{X} = \begin{bmatrix} \mathcal{H}_{11}\times_1\mathbf{u}_1^*\times_2\mathbf{v}_1 & \mathcal{H}_{12}\times_1\mathbf{u}_1^*\times_2\mathbf{v}_2 \\ \mathcal{H}_{21}\times_1\mathbf{u}_2^*\times_2\mathbf{v}_1 & \mathcal{H}_{22}\times_1\mathbf{u}_2^*\times_2\mathbf{v}_2 \end{bmatrix} \quad (8)$$

with notation similar to that for (4). The MIMO channel (8) remains a tensor channel similar to that in (3). The optimization of (8) should be similar to that for the matrix channel (4).

The ALS process may still be used by a matrix similar to (6). Given $\mathbf{v}_1$ and $\mathbf{v}_2$, we have $\mathbf{A}_{nm} = \mathcal{H}_{nm}\times_2\mathbf{v}_m$ with $n, m = 1, 2$. ALS algorithm can maximize $\|\mathbf{u}_1^H\mathbf{A}\mathbf{u}_2^*\|$ with $\mathbf{A} = \mathbf{A}_{11}\mathbf{A}_{22}^T - \mathbf{A}_{12}\mathbf{A}_{21}^T$. The ALS algorithm for tensor channel is given by Algorithm VI.

**ALGORITHM VI**: ALS FOR SPLIT MIMO TENSOR CHANNEL

Initial   Pick initial vectors $\mathbf{v}_1^{(0)}$ and $\mathbf{v}_2^{(0)}$, $k=0$.

Step 1   Increment $k$,
 a. Calculate $\mathbf{A}_{11}^{(k)} = \mathcal{H}_{11}\times_2\mathbf{v}_1^{(k-1)}$, $\mathbf{A}_{12}^{(k)} = \mathbf{H}_{12}\times_2\mathbf{v}_2^{(k-1)}$, $\mathbf{A}_{21}^{(k)} = \mathcal{H}_{21}\times_2\mathbf{v}_1^{(k-1)}$, $\mathbf{A}_{22}^{(k)} = \mathcal{H}_{22}\times_2\mathbf{v}_2^{(k-1)}$, and $\mathbf{A}^{(k)} = \mathbf{A}_{11}^{(k)}\mathbf{A}_{22}^{(k)T} - \mathbf{A}_{12}^{(k)}\mathbf{A}_{21}^{(k)T}$. Find $\mathbf{u}_1^{(k)}$ and $\mathbf{u}_2^{(k)}$ to minimize $\|\mathbf{A}^{(k)} - \lambda_1^{(k)}\mathbf{u}_1^{(k)}\mathbf{u}_2^{(k)T}\|^2$.
 b. Calculate $\mathbf{B}_{11}^{(k)} = \mathcal{H}_{11}\times_1\mathbf{u}_1^{(k)*}$, $\mathbf{B}_{12}^{(k)} = \mathcal{H}_{12}\times_1\mathbf{u}_1^{(k)*}$, $\mathbf{B}_{21}^{(k)} = \mathcal{H}_{21}\times_1\mathbf{u}_2^{(k)*}$, $\mathbf{B}_{22}^{(k)} = \mathcal{H}_{22}\times_1\mathbf{u}_2^{(k)*}$, and $\mathbf{B}^{(k)} = \mathbf{B}_{11}^{(k)T}\mathbf{B}_{22}^{(k)} - \mathbf{B}_{21}^{(k)T}\mathbf{B}_{12}^{(k)}$. Find $\mathbf{v}_1^{(k)}$ and $\mathbf{v}_2^{(k)}$ to minimize $\|\mathbf{B}^{(k)} - \lambda_1^{(k)}\mathbf{v}_1^{(k)*}\mathbf{v}_2^{(k)H}\|^2$.

Step 2   Repeat step 1 until the convergence of $\mathbf{u}_1^{(k)}$, $\mathbf{u}_2^{(k)}$, $\mathbf{v}_1^{(k)}$, and $\mathbf{v}_2^{(k)}$.

Step 3   Calculate $\mathcal{X}$ and obtain the corresponding singular values.

Because $\mathcal{X}$ (8) [also (3)] is still a tensor channel even after beamforming using the ALS Algorithm VI, the interpretation is not the same as the matrix channel $\Xi$ (4).

For each timing index $p$, there are two singular values for the matrix $\mathbf{X}_p$, giving us the singular values of $\sigma_{p,1}$ and $\sigma_{p,2}$, with $\sigma_{p,1} \geq \sigma_{p,2}$. The pre-coding matrix in the transmitter can be different in each time index $p$. Equivalently, for example, four different combinations of power of $\sigma_{1,n}^2 + \sigma_{2,m}^2$, $n, m = 1, 2$ and $m \neq n$, are possible if we just have two timing indices. Because the objective is to have two spatial channels with more or less the same power and thus the largest overall channel capacity, the optimal combinations in this example are $\sigma_{1,1}^2 + \sigma_{2,2}^2$ and $\sigma_{1,2}^2 + \sigma_{2,1}^2$ in which the larger singular value in one timing index is to combine with the smaller singular value in the other timing index.

In a greedy algorithm and for $2 \times 2$ MIMO, up to the timing index of $p$, the largest singular value should combine with the channel with less total power up to the previous timing index of $p - 1$. This may be generalized to $K$ spatial channels in which the singular values in ascent order are combined with the channels with power in descent order.

*D. Numerical Results*

Figure 6 shows the beamforming performance for $2 \times 2$ MIMO, tested using random Gaussian matrix or tensor channels. For matrix channel, the singular values are defined as the singular values for the channel matrix $\mathbf{H}$ and the singular values for $\Xi$ (4) in split MIMO architecture. For the same number of phase shifters, the channel matrix $\mathbf{H}$ is $16 \times 16$ in shared architecture and $32 \times 32$ in split architecture. The split architecture always has equal number of phase shifters, PAs, and antennas for each spatial stream.

For tensor matrix $\mathcal{H}$, number of transmitting and receiving antennas and antenna configuration are the same as the matrix channel. Number of timing index is two, the same as that in Figure 3. The singular values for Figure 6 are equivalently $\sqrt{\sigma_{1,1}^2 + \sigma_{2,2}^2}$ and $\sqrt{\sigma_{1,2}^2 + \sigma_{2,1}^2}$ for tensor channel, as described in Sec. IV.C.

The shared and split MIMO of Figure 6 for matrix channel are trained by a simple extension of the simple power method of Algorithm I and ALS Algorithm IV, respectively. The shared and split MIMO of Figure 6 for tensor channel is trained by Algorithm V and VI, respectively.

In MIMO beamforming, the tensor channel has more or less the same beamforming gain in both shared and split architectures. In split architecture has better performance than the shared architecture in the matrix channel.

## V. DISCUSSION

The numerical results of Figure 3 and Figure 6 are obtained by simulations. For random Gaussian matrix, the largest singular value is approximately $2\sqrt{N}$ for large $N$, consistent with the results of Figure 3 and Figure 6.

The singular values of Figure 3 do not seem consistent with the ideal beamforming gain of $20\log_{10}N + 10\log_{10}M$ from Sec. II.A in which all elements for $\mathbf{H}$ are deterministically the same instead of Gaussian random variable. The factor $20\log_{10}N$ assumes that all transmitters emit individually the same power as that in single antenna. However, the condition of $\|\mathbf{u}\| = 1$ assumes that the transmitter combined together emits the same power as that in single antenna. The factor of $10\log_{10}N$ can be added to Figure 3 to account for transmitter power emission.

In the shared MIMO architecture of Figure 4, each antenna emits a power on average twice that is in single antenna case of Figure 1 because the two transmitting beamforming vectors both have unity norm. In the split MIMO architecture of Figure 5, with twice number of antennas as in the numerical results of Figure 6, the overall output power should be the same as that in the corresponding shared MIMO architecture of Figure 4.

In practice but not discussed here, hybrid shared and split MIMO architecture is possible, for example, four independent data steams but two phase shifters per antenna.

Implementation issues are not discussed here. All algorithms here train the transmitting beamforming vector based on fixed receiving beamforming vector, and the other way around. This alternating optimization method should be consistent with the beam-search (or beam-refinement) protocols in both WirelessHD and IEEE 802.11ad. In all algorithms, calculation, $\tilde{\mathbf{u}}^{(k)} = \mathbf{H}\mathbf{v}^{(k-1)}$ in Algorithm I as an example, needs to be implemented based on channel measurement sequence. The signal processing for channel measurement is outside the scope of this paper.

Practical phase shifters may only provide constant amplitude phase shift. PA and LNA may also provide constant gain. The algorithms here can be modified accordingly.

## VI. CONCLUSION

For frequency-flat channels represented by channel matrix, the SISO beamforming based on analog phase shifters can be conducted using the classic simple power method. For frequency-selective channels represented by channel tensor, the SISO beamforming can be conducted using ALS method or HOPM.

For both matrix and tensor channels, beamforming can be performed for both shared and split MIMO architectures. MIMO beamforming is conducted iteratively to optimize the transmitting beamforming vectors by fixing receiving beamforming vectors, and then the other way around.

In the shared architecture, the MIMO beamforming algorithm is based on the extension of the power method or ALS method in SISO beamforming. In the split architecture, the algorithm is based on maximizing the product of singular values, equivalently the determinant of the $K \times K$ beamformed MIMO channel matrix. Beamforming algorithms for matrix and tensor channels can be derived using ALS method.

Ho *et al.*, MIMO beamforming in millimeter-wave directional Wi-Fi    8Ho *et al.*, MIMO beamforming in millimeter-wave directional Wi-Fi    8

Ho *et al.*, MIMO beamforming in millimeter-wave directional Wi-Fi 9

LIST OF FIGURES

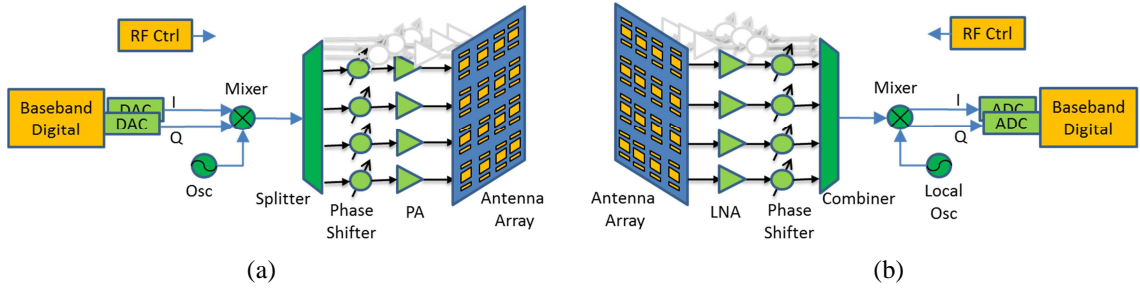

Figure 1 Schematic diagram of the 60-GHz millimeter-wave (a) transmitter and (b) receiver using antenna arrays.

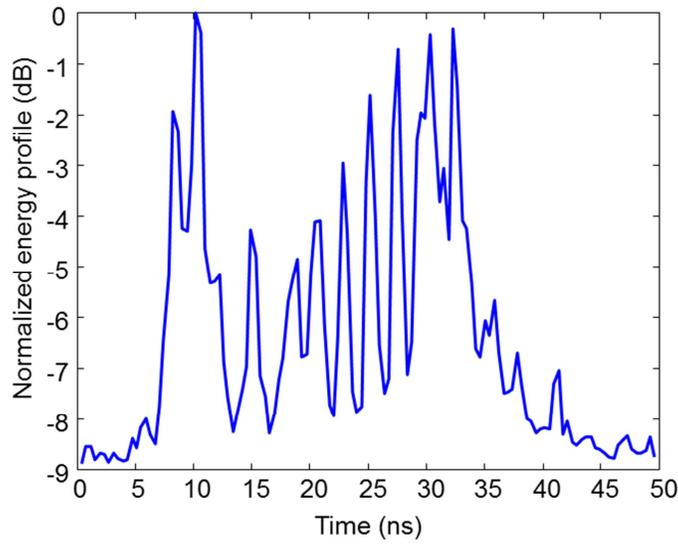

Figure 2 Energy profile for measured impulse responses between two antenna arrays. The time is relatively but not absolutely correct.

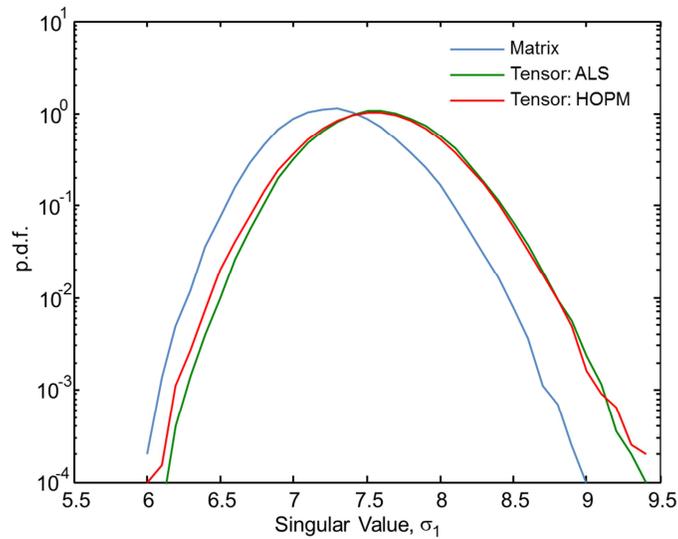

Figure 3 Distribution of singular value or its equivalent for SISO beamforming.

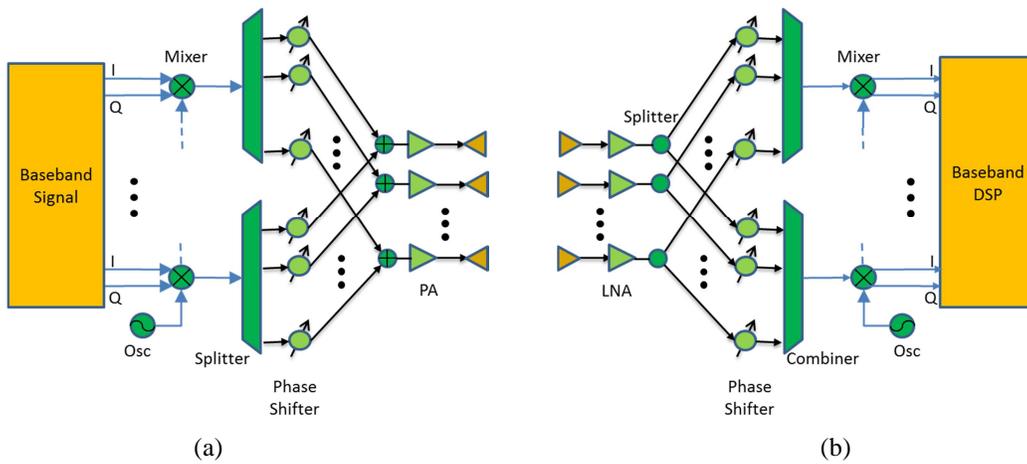

(a) (b)
Figure 4 Multiple phase shifters shared the same antenna in the shared MIMO architecture: (a) transmitter and (b) receiver.

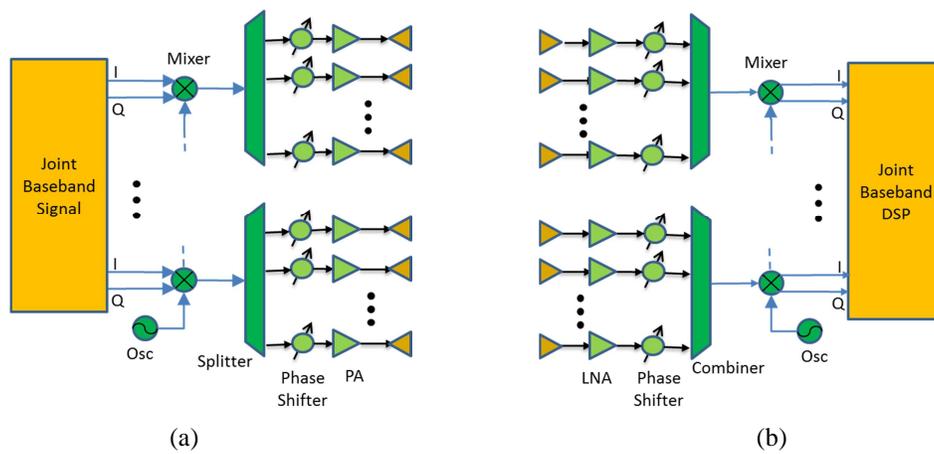

(a) (b)
Figure 5 Split MIMO architecture with different sets of antennas connecting to each data stream: (a) transmitter and (b) receiver.

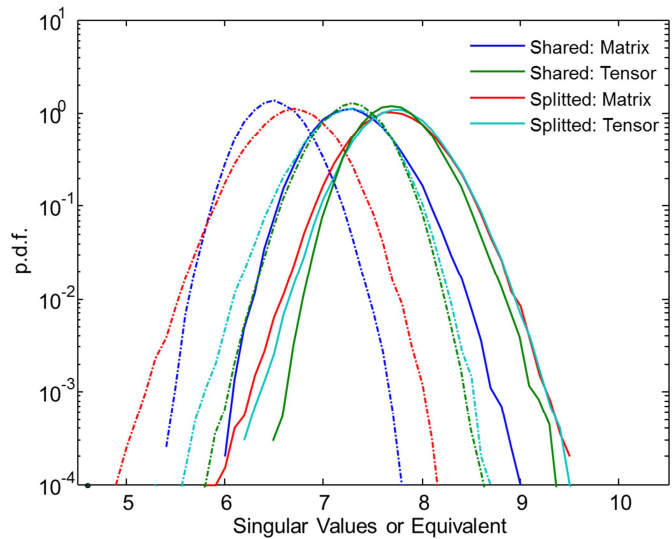

Figure 6 Distribution of singular values or their equivalent for $2 \times 2$ MIMO beamforming. Solid curves are for the stronger spatial channel. Dashed-dotted curves are for the weaker spatial channel. Curves with the same color are from the same MIMO simulation.